\documentclass[sigconf]{acmart}
\AtBeginDocument{%
  }

\newcommand{\name}{AVDA}

\usepackage{tabularx}
\usepackage{multirow}
\usepackage{booktabs}
\usepackage[utf8]{inputenc}
\usepackage[most]{tcolorbox}
\usepackage{xcolor}
\usepackage{lipsum} 
\usepackage{siunitx}    
\usepackage{tikz}
\usetikzlibrary{positioning, shapes.geometric, arrows.meta, calc}
\usepackage{amsmath}
\usepackage{url}

\tcbset{
  rqbox/.style={
    colback=gray!10,     
    colframe=gray!50,    
    boxrule=0.6pt,
    arc=4pt,             
    left=6pt, right=6pt, top=6pt, bottom=6pt,
    boxsep=4pt,
    breakable,
    enhanced,
    before skip=8pt,
    after skip=8pt
  }
}

\copyrightyear{2026}
\acmYear{2026}
\setcopyright{cc}
\setcctype{by-nc-nd}
\acmConference[FSE Companion '26]{34th ACM Joint European Software Engineering Conference and Symposium on the Foundations of Software Engineering}{July 05--09, 2026}{Montreal, QC, Canada}
\acmBooktitle{34th ACM Joint European Software Engineering Conference and Symposium on the Foundations of Software Engineering (FSE Companion '26), July 05--09, 2026, Montreal, QC, Canada}
\acmDOI{10.1145/3803437.3805261}
\acmISBN{/2026/07}

\begin{document}

\title{AVDA: Autonomous Vibe Detection Authoring for Cybersecurity}


\author{Fatih Bulut}
\affiliation{%
  \institution{Microsoft}
  \country{USA}}
\email{mbulut@microsoft.com}

\author{Carlo DePaolis}
\affiliation{%
  \institution{Microsoft}
  \country{USA}}
\email{mdepaolis@microsoft.com}

\author{Raghav Batta}
\affiliation{%
  \institution{Microsoft}
  \country{USA}}
\email{raghavbatta@microsoft.com}

\author{Anjali Mangal}
\affiliation{%
  \institution{Microsoft}
  \country{USA}}
\email{anjalimangal@microsoft.com}

\renewcommand{\shortauthors}{Bulut et al.}

\begin{abstract}
With the rapid advancement of AI in code generation, cybersecurity detection engineering faces new opportunities to automate traditionally manual processes. Detection authoring---the practice of creating executable logic that identifies malicious activities from security telemetry---is hindered by fragmented code across repositories, duplication, and limited organizational visibility. Current workflows remain heavily manual, constraining both coverage and velocity. In this paper, we introduce \name{}, a framework that leverages the Model Context Protocol (MCP) to automate detection authoring by integrating organizational context---existing detections, telemetry schemas, and style guides---into AI-assisted code generation. We evaluate three authoring strategies---Baseline, Sequential, and Agentic---across a diverse corpus of production detections and state-of-the-art LLMs. Our results show that Agentic workflows achieve a 19\% improvement in overall similarity score over Baseline approaches, while Sequential workflows attain 87\% of Agentic quality at 40$\times$ lower token cost. Generated detections excel at TTP matching (99.4\%) and syntax validity (95.9\%) but struggle with exclusion parity (8.9\%). Expert validation on a 22-detection subset confirms strong Spearman correlation between automated metrics and practitioner judgment ($\rho = 0.64$, $p < 0.002$). By integrating seamlessly into standard developer environments, \name{} provides a practical path toward AI-assisted detection engineering with quantified trade-offs between quality, cost, and latency.
\end{abstract}




\maketitle

\section{Introduction}

Artificial Intelligence (AI) and Machine Learning (ML) have long played a pivotal role in cybersecurity, enabling capabilities such as anomaly detection, entity extraction, and intrusion detection. Recent breakthroughs---particularly the emergence of Large Language Models (LLMs) and agentic workflows---have opened new avenues for innovation. Advances in code generation, exemplified by tools such as GitHub Copilot~\cite{github2021copilot}, OpenAI Codex~\cite{openai2021codex}, and Anthropic's Claude Code~\cite{anthropic2025claudecode}, have transformed software engineering practices by enabling automated, context-aware code synthesis.

The term vibe coding, coined by Andrej Karpathy in February 2025, describes a development style in which programmers express intent in natural language and rely on AI systems to generate implementation details through iterative, conversational interaction rather than manual coding. \name{} extends this paradigm to cybersecurity detection engineering: authors describe threats in natural language and leverage AI-assisted workflows to produce executable detection logic, hence ``Vibe Detection Authoring.'' 

Detection engineering, a cornerstone of modern cybersecurity, has evolved from ad hoc rule creation into a structured discipline that mirrors the software development lifecycle. Contemporary detection-as-code practices \cite{splunkdac2024} emphasize version control, testing, and continuous integration, aligning detection development with established engineering principles. Detection artifacts vary widely in language, format, and complexity---from simple atomic rules expressed in standardized query languages to sophisticated behavioral analytics. Critically, detection authoring is fundamentally a code generation task: authors translate threat concepts into executable logic tailored to specific platforms and telemetry schemas. This makes detection engineering a natural candidate for LLM-assisted automation.

Despite the maturation of detection-as-code practices, detection development remains challenging. As enterprise attack surfaces expand through cloud adoption and heterogeneous security ecosystems, detection workflows often become fragmented across multiple repositories and platforms. This fragmentation introduces duplication, inconsistency, and significant manual overhead, ultimately limiting both visibility and coverage. Traditional authoring workflows---identifying protection gaps, drafting detection logic, performing unit tests, and validating against real-world data---are resource-intensive and do not scale with the velocity of modern threats. In large enterprises operating across multiple vendors, overlapping rules and blind spots are common, increasing maintenance burden and reducing organizational resilience.

The advent of state-of-the-art LLMs presents an opportunity to address these challenges. Recent innovations such as the Model Context Protocol (MCP)~\cite{mcp2024} enable seamless context sharing across tools and environments, facilitating interoperability between AI assistants and organizational knowledge bases. MCP allows detection authors to query existing detection portfolios, retrieve telemetry schemas, and generate platform-specific code---all within a unified interface. Coupled with widespread adoption of developer tools such as Visual Studio Code, these advancements create an opportunity to streamline detection authoring through AI-driven automation.

In this paper, we introduce \name{}, a framework designed to automate detection authoring workflows by integrating LLM-based code generation with organizational context. \name{} leverages MCP to provide detection authors with access to existing detections, schema information, and platform-specific style guides during the authoring process. By embedding AI assistance directly into developer environments, \name{} reduces duplication, improves consistency, and accelerates the development of high-quality detections while preserving human oversight.

We evaluate \name{} across three authoring strategies---Baseline (zero-shot generation), Sequential (retrieval-augmented generation), and Agentic (iterative tool-orchestrated reasoning)---using 92 production detections spanning 5 platforms and 3 languages. Our evaluation encompasses 11 models and 21 configurations across multiple LLM families, including reasoning-capable models at varying reasoning effort levels, producing 5,796 generated detection artifacts. Results demonstrate that Agentic workflows achieve the highest overall similarity score (mean 0.447), representing a 19\% improvement over Baseline approaches. Sequential workflows offer a practical alternative, achieving 87\% of Agentic quality at 40$\times$ lower token cost. Expert validation on a 22-detection subset confirms strong correlation between automated metrics and practitioner judgment (Spearman $\rho = 0.64$, $p < 0.002$), establishing the reliability of our evaluation framework.

Our contributions are as follows:
\begin{itemize}
    \item \textbf{\name{} Framework:} A retrieval-augmented detection authoring system that leverages the Model Context Protocol (MCP) to integrate organizational context---including existing detections, telemetry schemas, and style guides---into AI-assisted code generation within standard developer tools.
    
    \item \textbf{Comprehensive Empirical Evaluation:} A systematic comparison of Baseline, Sequential, and Agentic workflows across 11 models, 21 configurations, and 92 production detections spanning 5 platforms and 3 languages, quantifying trade-offs between quality (19\% improvement for Agentic) and cost (40$\times$ token overhead). We identify that generated detections excel at TTP matching (99.4\%) and syntax validity (95.9\%) but struggle with exclusion parity (8.9\%).
    
    \item \textbf{Validated Evaluation Methodology:} An automated evaluation framework combining LLM-as-a-judge semantic assessment with embedding-based similarity, validated against expert ratings on a 22-detection subset with statistically significant correlation ($\rho = 0.64$, $p < 0.002$), enabling scalable quality assessment for AI-generated detections.
    
    \item \textbf{Practical Guidance for Practitioners:} Empirical insights on model selection (reasoning-capable models outperform), workflow trade-offs (Sequential as cost-effective alternative), and failure modes (schema hallucination, missing exclusion logic), informing adoption of AI-assisted detection authoring in enterprise settings.
\end{itemize}

The remainder of this paper is organized as follows. Section~\ref{sec:related} reviews related work, providing context for existing approaches and their limitations. Section~\ref{sec:system} describes the system design of \name{}, detailing its architecture and integration with developer tools. Section~\ref{sec:evaluation} presents our evaluation methodology and results. Section~\ref{sec:discussion} discusses key findings, limitations, and lessons learned. Finally, Section~\ref{sec:conclusion} concludes the paper and outlines directions for future research.
\section{Related Work}
\label{sec:related}

Recent advances in large language models have prompted exploration of AI-assisted security detection authoring. Schwartz et al.~\cite{schwartz2024llmcloudhunterharnessingllmsautomated} describe a system that extracts generic-signature detection rules from cloud-focused cyber threat intelligence (CTI) reports, converting unstructured CTI into sigma rules and Splunk queries via LLM pipeline. In contrast, \name{} targets platform-specific detection authoring across heterogeneous environments (KQL, Python, Scala), compares three increasingly autonomous workflows (baseline, sequential, agentic), and evaluates detection quality through ten binary semantic criteria enabling diagnostic analysis of individual failure modes. RuleGenie~\cite{shukla2025rulegeniesiemdetectionrule} has explored LLM-aided optimization of existing detection rule sets. It uses embeddings and chain-of-thought reasoning to identify redundant SIEM rules and recommend consolidations. While complementary to our generation-focused approach, such optimization would be applied after \name{} produces candidate detections. 
GRIDAI~\cite{li2025gridaigeneratingrepairingintrusion} proposes a multi-agent LLM framework for generating \emph{and repairing} network intrusion detection rules. Their system comprises multiple agents. While GRIDAI addresses network-layer intrusion detection from traffic samples, our work targets any type of detection authoring from natural language descriptions. Similar to \cite{shukla2025rulegeniesiemdetectionrule}, GRIDAI's key contribution is reducing ruleset redundancy by repairing existing rules for attack variants; in contrast, \name{} focuses on comparing workflow automation levels
(baseline, sequential RAG, and agentic) to determine optimal LLM orchestration strategies for enterprise security platforms. The authors in~\cite{muzammil2025smalllanguagemodelssecurity} investigate small language models for KQL query generation in SOC workflows, similar to \cite{tang2025nl2kqlnaturallanguagekusto}, demonstrating that fine-tuned models can match larger models on narrow domains, though their evaluation focuses on query correctness rather than comprehensive detection quality metrics. While their work targets \emph{query generation} for log analysis and threat hunting, \name{} addresses \emph{detection authoring}---generating executable rules that run
continuously in production environments. Both systems share a schema-aware retrieval component, but \name{} explores varying \emph{autonomy levels} (baseline, sequential,
agentic) and generates complex multi-language detections (KQL, Python/PySpark, Scala) rather than KQL queries alone.

The Detection as Code (DaC) movement applies software engineering practices to detection lifecycle management~\cite{splunkdac2024}. Industry surveys indicate that while 63\% of security teams aspire to adopt DaC practices, only 35\% have implemented them, citing challenges in version control, automated testing, and environment drift. Palantir's Alerting and Detection Strategy (ADS) framework~\cite{palantirads2017} 
exemplifies mature DaC practices with structured documentation, mandatory peer 
review, and version-controlled pipelines. 

Sublime Security's Automated Detection Engineer (AD\'{E})~\cite{sublimeade2024, bertiger2025evaluatingllmgenerateddetection} represents an industry deployment of LLM-driven detection authoring, evaluating generated rules across detection accuracy, robustness, and economic cost, with production deployments reporting high precision comparable to human-authored rules. Bertiger et al.~\cite{bertiger2025evaluatingllmgenerateddetection} present an evaluation framework introducing a holdout-based methodology that compares LLM-generated rules against human-authored baselines using expert-inspired metrics. NL2KQL~\cite{tang2025nl2kqlnaturallanguagekusto} provides a benchmark for security log query generation evaluating both syntactic and semantic correctness, but benchmarks using synthetic datasets may not reflect production complexity.

The Model Context Protocol (MCP)~\cite{mcp2024} provides an open standard for connecting AI applications to external systems, enabling tool-augmented LLM workflows by standardizing data source access and multi-turn context retrieval. Recent work distinguishes between \emph{sequential} workflows that chain tool invocations without dynamic adaptation and \emph{agentic} systems that incorporate feedback loops for iterative refinement~\cite{anthropic2025agents}. Structured workflows often outperform autonomous agents on deterministic tasks, while agents excel when problems require exploration or error recovery---detection authoring exhibits characteristics of both.

\name{} differentiates itself from prior work by combining: (1) support for heterogeneous enterprise environments spanning multiple languages (Python/PySpark, KQL, Scala) and multiple platforms; (2) MCP-grounded tool access for schema validation and telemetry introspection; (3) a systematic comparison of Baseline, Sequential, and Agentic workflows across 11 state-of-the-art models and reasoning configurations; and (4) rigorous evaluation against 92 production detections validated by expert correlation.
\section{System Design}
\label{sec:system}

\begin{figure*}[htbp]
  \centering
  \includegraphics[width=0.85\textwidth]{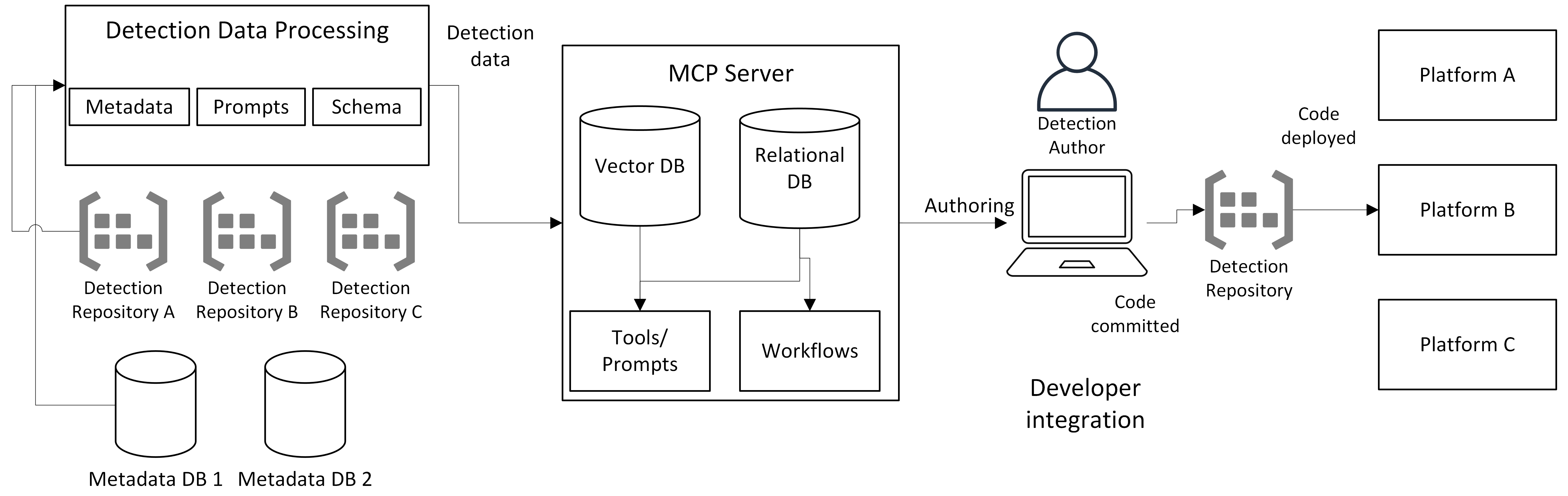}
  \caption{\name{} architecture. Detection artifacts flow from organizational repositories through the Data Processing layer, which populates vector and relational stores. The MCP Server exposes these assets to LLM-powered authoring workflows via standardized tools. Detection authors interact through IDE extensions or CLI, with generated code following standard DevOps pipelines to deployment.}
  \label{fig:architecture}
\end{figure*}

Figure~\ref{fig:architecture} illustrates the \name{} architecture, which comprises four main components: (1) a \emph{Detection Data Processing} layer that ingests and normalizes detection artifacts from organizational repositories; (2) an \emph{MCP Server} that exposes tools for semantic search, schema introspection, and code generation; (3) \emph{Authoring Workflows} that orchestrate LLM interactions in baseline, sequential, or agentic modes; and (4) \emph{Developer Integration} that embeds these capabilities into standard IDE and CI/CD environments. The design prioritizes three objectives: \emph{scalability} across large detection portfolios, \emph{interoperability} across heterogeneous platforms, and \emph{automation} of repetitive authoring tasks while preserving human oversight.

\subsection{Detection Data Processing}
\label{subsec:data-processing}

This layer ingests and normalizes detection artifacts from heterogeneous sources. It processes two primary inputs: (i) detection code from organizational repositories, treated as the canonical source of truth, and (ii) metadata from wikis and platform catalogs, which is often incomplete or inconsistent.

To standardize these artifacts, the system applies platform-specific parsers augmented by LLM-based canonicalization for schema alignment. A unified schema captures key fields including:
\begin{itemize}
    \item \texttt{detection\_id}, \texttt{name}, \texttt{description} — unique identifier and documentation
    \item \texttt{platform}, \texttt{language} — execution environment (e.g., XDR/PySpark, Sentinel/KQL)
    \item \texttt{mitre\_tactics}, \texttt{mitre\_techniques} — ATT\&CK mappings for coverage analysis and search
    \item \texttt{data\_sources} — telemetry tables and signals consumed
    \item \texttt{original\_content} — raw detection code for retrieval and comparison
    \item \texttt{repository}, \texttt{file\_path} — provenance for lineage tracking
\end{itemize}
Missing metadata (e.g., MITRE mappings) is inferred from code analysis and related documentation.

The system also auto-generates two assets consumed by the MCP server: (i) \emph{prompt templates} encoding platform conventions and style guides, and (ii) \emph{context schemas} summarizing available tables, columns, and join patterns derived from existing detections.

Outputs are persisted in two complementary stores: a \emph{vector database} (using OpenAI's \texttt{text-embedding-3-large}) for semantic search and similarity queries, and a \emph{relational database} (SQLite) for structured metadata, lineage tracking, and audit operations. This dual-store architecture supports both exploratory retrieval and deterministic lookups required by the MCP tools.

\subsection{MCP Server and Tool Suite}
\label{subsec:mcp-core}

The Model Context Protocol (MCP)~\cite{mcp2024} provides a standardized interface for LLM agents to interact with external tools and data sources. \name{} implements an MCP server that exposes organizational detection knowledge through three abstraction types: \emph{tools} (executable functions), \emph{resources} (data access), and \emph{prompts} (reusable templates).

\begin{table}[t]
\centering
\caption{MCP tools exposed by \name{}.}
\label{tab:mcp_tools}
\renewcommand{\arraystretch}{1.1}
\footnotesize
\begin{tabular}{@{} l l p{3.5cm} @{}}
\toprule
\textbf{Category} & \textbf{Tool} & \textbf{Description} \\
\midrule
\multirow{4}{*}{Retrieval} 
  & \texttt{semantic\_search} & Vector similarity search over detection embeddings \\
  & \texttt{search\_by\_mitre} & Retrieve detections by MITRE ATT\&CK technique or tactic \\
  & \texttt{search\_by\_platform} & Filter detections by platform and language \\
  & \texttt{get\_content} & Fetch full source code and metadata by ID \\
\midrule
\multirow{4}{*}{Schema} 
  & \texttt{get\_telemetry\_fields} & List fields for a telemetry table \\
  & \texttt{get\_supported\_actions} & Return actions for a telemetry table \\
  & \texttt{get\_actions\_and\_tables} & Enumerate action--table mappings \\
  & \texttt{get\_best\_table} & Recommend table for a detection action \\
\midrule
\multirow{2}{*}{Similarity} 
  & \texttt{get\_similar} & Find detections by cosine similarity \\
  & \texttt{get\_details} & Retrieve extended metadata and mappings \\
\bottomrule
\end{tabular}
\end{table}

Table~\ref{tab:mcp_tools} summarizes the tools exposed by \name{}, organized into three categories:

\begin{itemize}
    \item \textbf{Retrieval Tools:} Enable semantic and structured search across the detection portfolio, supporting filters by platform, MITRE technique, and language.
    \item \textbf{Schema Tools:} Provide access to telemetry schemas, including available tables, fields, and supported detection actions for each platform.
    \item \textbf{Similarity Tools:} Identify related detections using vector similarity, supporting deduplication and exemplar-based generation.
\end{itemize}

This tool suite represents the \emph{minimal foundation} used in our evaluations. The MCP architecture is intentionally extensible, allowing organizations to add domain-specific tools as detection engineering needs evolve. Examples of additional tools that could enhance the workflow include: querying IOC feeds and threat actor profiles to enrich detection context; executing queries against sample datasets or replay logs to validate logic before deployment; identifying gaps in MITRE ATT\&CK coverage or overlapping detections across the portfolio; integrating with CI/CD systems to lint, test, and promote detections through staging environments; and retrieving historical alert data to inform tuning decisions and exclusion logic.

\subsection{Authoring Workflows}
\label{subsec:workflows}

\name{} supports three authoring workflows that represent increasing levels of tool integration and autonomy. These workflows are central to our evaluation (Section~\ref{sec:evaluation}).

\textbf{Baseline.} The simplest approach provides the LLM with a detection specification (target TTP, platform, language and natural language description) without access to MCP tools. This zero-shot generation serves as a lower bound, isolating the model's intrinsic capability from retrieval augmentation.

\textbf{Sequential.} This workflow follows a deterministic pipeline that mirrors structured human practice:
\begin{enumerate}
    \item \emph{Extract} detection metadata (MITRE techniques, platform, keywords) via LLM preprocessing
    \item \emph{Query} the vector database via \texttt{semantic\_search} to retrieve the top-10 similar detections
    \item \emph{Retrieve} full content for the top-3 results via \texttt{get\_content}
    \item \emph{Generate} detection code using retrieved exemplars as grounding context
\end{enumerate}
The sequential approach provides retrieval-augmented generation (RAG) with predictable cost and latency.

\textbf{Agentic.} This workflow adopts a ReAct-style paradigm~\cite{yao2023reactsynergizingreasoningacting} where the LLM iteratively reasons, invokes tools, observes results, and refines its approach. At each iteration, the agent receives the full conversation history---including prior tool calls and their outputs---and autonomously decides which tool to invoke next. The agent may call \texttt{semantic\_search} to find similar detections, retrieve their content via \texttt{get\_content}, query telemetry schemas with \texttt{get\_telemetry\_fields}, or explore MITRE mappings---all without predetermined ordering. This exploratory loop continues until the agent determines it has sufficient context to generate the final detection code, or until reaching a maximum of 150 iterations. While more capable for complex or ambiguous detection requirements, this approach exhibits higher token consumption and latency variability due to the multi-turn reasoning overhead.

\begin{figure*}[t]
\centering
\begin{tikzpicture}[
    node distance=0.6cm,
    box/.style={rectangle, draw, rounded corners, minimum width=1.4cm, minimum height=0.6cm, align=center, font=\tiny},
    promptbox/.style={box, fill=blue!15},
    llmbox/.style={box, fill=orange!20},
    mcpbox/.style={box, fill=green!20},
    outputbox/.style={box, fill=purple!15},
    databox/.style={box, fill=gray!15, dashed},
    decisionbox/.style={diamond, draw, fill=yellow!20, minimum width=0.7cm, minimum height=0.6cm, align=center, font=\tiny, inner sep=1pt},
    arrow/.style={->, >=stealth, thick},
    looparrow/.style={->, >=stealth, thick, dashed},
    rowlabel/.style={font=\bfseries\small, align=left}
]

\def\rowA{0}       
\def\rowB{-2.2}    
\def\rowC{-4.8}    

\draw[gray, dashed] (-1.5, -1.1) -- (14, -1.1);
\draw[gray, dashed] (-1.5, -3.4) -- (14, -3.4);

\node[rowlabel, anchor=east] at (-0.3, \rowA) {Baseline};
\node[rowlabel, anchor=east] at (-0.3, \rowB) {Sequential};
\node[rowlabel, anchor=east] at (-0.3, \rowC) {Agentic};

\node[promptbox] (b-prompt) at (0.8, \rowA) {Detection\\Desc.};
\node[llmbox] (b-llm) at (3, \rowA) {LLM\\Generate};
\node[outputbox] (b-output) at (5.2, \rowA) {Detection\\Code};

\draw[arrow] (b-prompt) -- (b-llm);
\draw[arrow] (b-llm) -- (b-output);

\node[font=\tiny, text=gray, align=left] at (7.5, \rowA) {Single call, no tools};

\node[promptbox] (s-prompt) at (0.8, \rowB) {Detection\\Desc.};
\node[llmbox] (s-extract) at (2.6, \rowB) {Extract\\Keywords};
\node[mcpbox] (s-search) at (4.4, \rowB) {Semantic\\Search};
\node[databox] (s-topk) at (6.2, \rowB) {Top-10};
\node[mcpbox] (s-retrieve) at (8, \rowB) {Get\\Content};
\node[databox] (s-examples) at (9.8, \rowB) {Top-3\\Examples};
\node[llmbox] (s-generate) at (11.6, \rowB) {Generate\\w/ Context};
\node[outputbox] (s-output) at (13.4, \rowB) {Detection\\Code};

\draw[arrow] (s-prompt) -- (s-extract);
\draw[arrow] (s-extract) -- (s-search);
\draw[arrow] (s-search) -- (s-topk);
\draw[arrow] (s-topk) -- (s-retrieve);
\draw[arrow] (s-retrieve) -- (s-examples);
\draw[arrow] (s-examples) -- (s-generate);
\draw[arrow] (s-generate) -- (s-output);

\node[promptbox] (a-prompt) at (0.8, \rowC) {Detection\\Desc.};
\node[llmbox] (a-agent) at (2.8, \rowC) {Agent\\(Reason)};
\node[decisionbox] (a-decision) at (4.8, \rowC) {Tool?};
\node[mcpbox] (a-tools) at (6.8, \rowC) {MCP\\Tools};
\node[databox] (a-observe) at (8.8, \rowC) {Observe};
\node[decisionbox] (a-done) at (10.8, \rowC) {Done?};
\node[outputbox] (a-output) at (13, \rowC) {Detection\\Code};

\draw[arrow] (a-prompt) -- (a-agent);
\draw[arrow] (a-agent) -- (a-decision);
\draw[arrow] (a-decision) -- (a-tools) node[midway, above, font=\tiny] {yes};
\draw[arrow] (a-tools) -- (a-observe);
\draw[arrow] (a-observe) -- (a-done);
\draw[arrow] (a-done) -- (a-output) node[midway, above, font=\tiny] {yes};

\draw[arrow] (a-decision.south) -- ++(0, -0.6) -| (a-done.south) node[pos=0.05, below, font=\tiny] {no};

\draw[looparrow] (a-done.north) -- ++(0, 0.7) -| (a-agent.north) node[pos=0.02, above, font=\tiny] {no (max 150)};

\end{tikzpicture}
\caption{Comparison of detection authoring workflows.}
\label{fig:workflows}
\end{figure*}
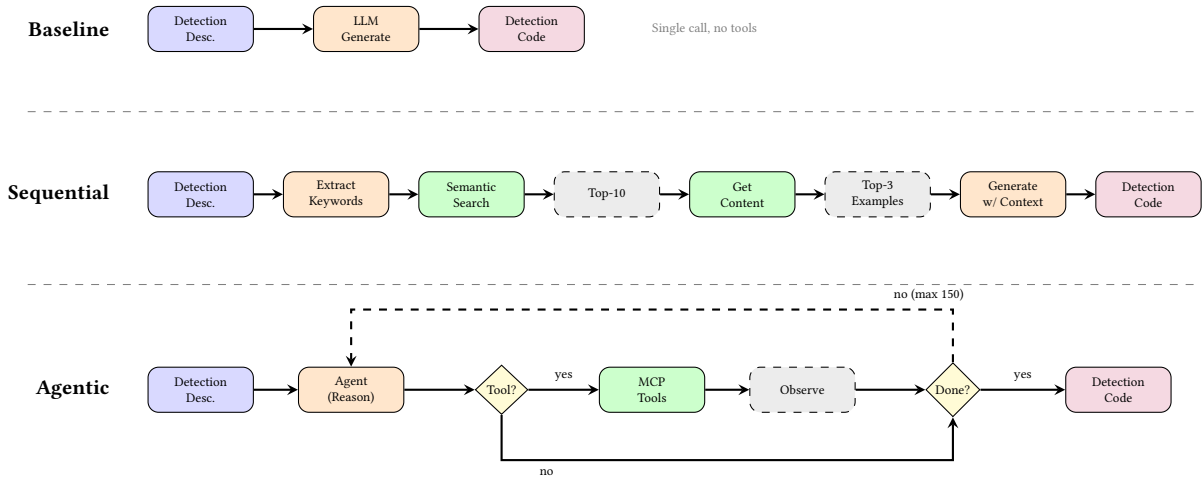

Figure~\ref{fig:workflows} illustrates the three workflows. The choice of workflow involves trade-offs between quality, cost, and predictability that we quantify in Section~\ref{sec:evaluation}.

\subsection{Developer Integration}
\label{subsec:dev-integration}

\name{} integrates with developer workflows through multiple entry points. The primary interface is an MCP-enabled extension for Visual Studio Code, allowing authors to invoke tools, generate detections, and validate outputs without leaving the editor's chat window. This integration leverages the MCP client capabilities built into modern AI coding assistants.

Beyond the IDE, \name{} aligns with standard DevOps practices. Generated detections follow the conventional pipeline: \emph{edit} $\rightarrow$ \emph{commit} $\rightarrow$ \emph{pull request} $\rightarrow$ \emph{CI checks} $\rightarrow$ \emph{review} $\rightarrow$ \emph{deploy}. MCP tools can programmatically create pull requests with structured summaries, reducing review overhead while preserving human approval gates.

A continuous improvement loop complements authoring. Feedback from automated checks (syntax validation, schema compliance) and human reviews is captured and incorporated into prompt templates, enabling iterative refinement of generation quality over time.
\section{Evaluation}
\label{sec:evaluation}

We evaluate \name{} across multiple dimensions: detection code quality, workflow efficiency, and model performance characteristics. Our primary goal is to assess how closely generated detections align with ground-truth implementations in terms of both semantic fidelity (functional correctness and intended behavior) and syntactic similarity (structural correspondence). To enable reproducible evaluation at scale, \name{} employs an automated composite metric that combines LLM-based semantic judgment with embedding similarity and surface-level syntactic measures; we detail this metric and validate it against expert ratings in \S\ref{subsec:metrics} and RQ1. Specifically, we address the following research questions:

\begin{itemize}
    \item \textbf{RQ1 (Expert Validation):} How well do automated metrics align with security practitioner ratings?
    \item \textbf{RQ2 (Workflow Comparison):} What are the performance differences among Baseline, Sequential, and Agentic approaches?
    \item \textbf{RQ3 (Criterion Analysis):} Which LLM-as-a-Judge criteria have the highest and lowest pass rates, and how does tool-assisted retrieval affect them?
    \item \textbf{RQ4 (Reasoning Effort):} How do high, medium, and low reasoning tiers compare?
    \item \textbf{RQ5 (Top-5 Leaderboard):} What model--approach combinations occupy the Top-5 positions?
    \item \textbf{RQ6 (Token Efficiency):} How do workflows compare in token consumption?
    \item \textbf{RQ7 (Platform \& Language):} How do scores vary across platforms and languages?
    \item \textbf{RQ8 (Model Timeline):} Do newer model releases yield better detection quality?
\end{itemize}

\subsection{Experimental Setup}
\label{subsec:exp-setup}

We evaluate \name{} on a curated corpus of 92 production detections sourced from five Microsoft detection platforms: two public platforms (Microsoft Sentinel and Microsoft Defender XDR) and three internal security platforms (referred to as Internal Platforms A, B, and C in subsequent tables to preserve confidentiality). The corpus spans three languages---PySpark (Python) (43\%), KQL (35\%), and Scala (22\%)---reflecting the heterogeneous tooling found in enterprise security operations. Detections are approximately balanced across platforms (20 per platform), with XDR contributing 12 detections.

All evaluation detections were strictly withheld from the MCP server's retrieval indices to eliminate data leakage. Although there may be publicly available documentation for Sentinel and XDR that may have influenced pretraining of frontier models, none of the 92 evaluation detections were exposed during generation via MCP tools, ensuring a clean separation between training and test conditions.

\begin{table}[t]
\centering
\caption{Evaluation dataset summary.}
\label{tab:dataset-summary}
\small
\begin{tabular}{@{} l r @{}}
\toprule
\textbf{Characteristic} & \textbf{Value} \\
\midrule
Total detections & 92 \\
Platforms & 5 \\
Languages & 3 \\
MITRE tactics covered & 14 \\
MITRE techniques covered & 66 \\
Lines of code (range) & 12--3{,}280 \\
Lines of code (median) & 94 \\
Expert-validated subset & 22 \\
\bottomrule
\end{tabular}
\end{table}

Table~\ref{tab:dataset-summary} summarizes the evaluation corpus. The selected detections span all 14 MITRE ATT\&CK \cite{mitre_attack} tactics and 66 unique techniques. Complexity varies---from simple 12-line to 3{,}280-line multi-stage detection pipelines (mean: 286 LoC)---ensuring the evaluation captures a broad spectrum of authoring challenges representative of enterprise-scale detection engineering.

\name{}'s preprocessing pipeline ingested 10{,}127 detection artifacts from 9 different detection platforms to construct a rich contextual foundation comprising schemas, exemplars, and platform-specific style guides. The 92 evaluation detections were explicitly held out from this corpus to prevent data leakage. For expert validation (\S\ref{subsec:results}, RQ1), we selected a stratified subset of 22 detections spanning 13 of the 14 MITRE ATT\&CK tactics to ensure diverse coverage.

We compare three authoring workflows representing increasing levels of sophistication and autonomy: baseline, sequential, and agentic. To ensure prompt fairness, platform-specific prompt styles encoded as MCP prompts are excluded from our evaluation. Additionally, only two of the five platforms (XDR and Platform B) expose schema information through MCP tools, enabling us to evaluate the effectiveness of including runtime schema access in the agentic workflow.

\subsection{Metrics}
\label{subsec:metrics}

Our evaluation framework centers on two complementary dimensions: \emph{semantic similarity}, which captures functional correctness, and \emph{syntactic similarity}, which measures structural resemblance. Together, these dimensions provide a holistic view of detection quality.

\paragraph{Semantic Similarity.}
We employ two complementary approaches. First, an \emph{LLM-as-a-judge} protocol uses GPT-4.1 (temperature $\tau=0$) to assess functional equivalence between generated and gold-standard detections \cite{zheng2023judgingllmasajudgemtbenchchatbot}. Rather than soliciting a single holistic score---which can be unreliable---we decompose evaluation into ten binary criteria (Table~\ref{tab:eval_criteria}), each answered True/False. The LLM-judge score is computed as:
\begin{equation}
    S_{\text{LLM}} = \frac{1}{n} \sum_{i=1}^{n} c_i, \quad c_i \in \{0, 1\}
\end{equation}
where $n=10$ is the number of criteria and $c_i$ indicates whether criterion $i$ is satisfied.

Second, we compute \emph{embedding similarity} using cosine similarity between vector representations from OpenAI's text-embedding-3-large model:
\begin{equation}
    S_{\text{embed}} = \frac{\mathbf{v}_g \cdot \mathbf{v}_r}{\|\mathbf{v}_g\| \|\mathbf{v}_r\|}
\end{equation}
where $\mathbf{v}_g$ and $\mathbf{v}_r$ are the embedding vectors of the generated and reference detections, respectively.

\begin{table}[h!]
\centering
\caption{Evaluation Questions for LLM-as-a-Judge}
\label{tab:eval_criteria}
\footnotesize
\renewcommand{\arraystretch}{1.15}
\begin{tabularx}{\columnwidth}{@{} c X @{}}
\toprule
\textbf{Question} & \textbf{Description} \\
\midrule
Q1 (ttp\_match) & Does the candidate detection target the exact same TTP as the Gold detection? \\
\addlinespace
Q2 (logic\_equivalence) & Is the candidate's core detection logic (filters, joins, aggregations) functionally equivalent to the Gold detection? \\
\addlinespace
Q3 (schema\_accuracy) & Does the candidate use the correct tables/fields without hallucinations? \\
\addlinespace
Q4 (syntax\_validity) & Is the candidate syntactically valid for the target language/platform? \\
\addlinespace
Q5 (indicator\_alignment) & Does the candidate check for the same key indicators/signals as Gold? \\
\addlinespace
Q6 (exclusion\_parity) & Does the candidate implement the same (or stronger) exclusion logic to control false positives? \\
\addlinespace
Q7 (robustness) & Is the candidate equally resilient to common evasions (case, encoding, minor variations)? \\
\addlinespace
Q8 (data\_source\_correct) & Does the candidate pull from the correct data source for this detection? \\
\addlinespace
Q9 (output\_alignment) & Does the candidate emit results to the expected destination (e.g., Alerts table) with required fields? \\
\addlinespace
Q10 (library\_usage) & Does the candidate leverage the correct platform-specific libraries/functions required for this detection? \\
\bottomrule
\end{tabularx}
\end{table}

\paragraph{Syntactic Similarity.}
We evaluate structural resemblance using two surface-level metrics. First, \emph{ROUGE-L F1} measures the longest common subsequence (LCS) between generated and reference code:
\begin{equation}
    S_{\text{ROUGE-L}} = \frac{(1 + \beta^2) \cdot P_{\text{lcs}} \cdot R_{\text{lcs}}}{R_{\text{lcs}} + \beta^2 \cdot P_{\text{lcs}}}, \quad P_{\text{lcs}} = \frac{|\text{LCS}|}{|g|}, \quad R_{\text{lcs}} = \frac{|\text{LCS}|}{|r|}
\end{equation}
where $|g|$ and $|r|$ are the lengths of the generated and reference sequences, and $\beta=1$ for balanced F1.

Second, \emph{normalized Levenshtein similarity} quantifies edit distance:
\begin{equation}
    S_{\text{Lev}} = 1 - \frac{\text{Lev}(g, r)}{\max(|g|, |r|)}
\end{equation}
where $\text{Lev}(g, r)$ is the minimum number of single-character edits (insertions, deletions, substitutions) to transform $g$ into $r$ \cite{levens-1, levens-2}. These metrics capture surface-level similarity without implying functional correctness---important because semantically equivalent detections may differ substantially in formatting, variable naming, or stylistic conventions.

\paragraph{Composite Scores.}
We aggregate individual metrics into composite scores:
\begin{align}
    S_{\text{syntactic}} &= 0.5 \times S_{\text{ROUGE-L}} + 0.5 \times S_{\text{Lev}} \\
    S_{\text{semantic}} &= 0.8 \times S_{\text{LLM}} + 0.2 \times S_{\text{embed}} \\
    S_{\text{overall}} &= 0.8 \times S_{\text{semantic}} + 0.2 \times S_{\text{syntactic}}
\end{align}

The syntactic score equally weights ROUGE-L and Levenshtein as both measure surface-level similarity with no strong prior for one over the other. Within the semantic score, we weight the LLM-judge at 0.8 because its criterion-based decomposition directly evaluates functional properties (TTP matching, schema accuracy, logic equivalence) and enables diagnostic analysis of failure modes, whereas embedding similarity provides a complementary but less interpretable holistic signal. The overall 80/20 semantic-syntactic weighting prioritizes functional behavior over textual resemblance; we validate this choice empirically in RQ1.

\paragraph{Scope and Limitations.}
Our evaluation measures alignment with human-authored gold-standard detections \emph{without runtime execution}. This design enables scalable evaluation across 92 detections with different model configurations and workflows, but means scores reflect authoring proximity rather than verified correctness. In practice, generated detections serve as high-fidelity starting points that engineers refine---a workflow validated by expert assessment (RQ1).

\subsection{Models}
\label{subsec:models}

We evaluate 11 models spanning 21 configurations: 5 chat/completion models (GPT-4.1, GPT-4.1-mini, GPT-4o, GPT-4o-mini, GPT-5.1-chat) and 6 reasoning models (GPT-5, GPT-5-mini, GPT-5.1, o1, o3, o3-pro). Chat models use their default configuration, while reasoning models are tested at three reasoning-effort levels corresponding to the API  parameter: low (minimal chain-of-thought), medium (balanced reasoning depth), and high (extended thinking); o3-pro supports only high.

All generations use temperature $\tau = 0$ for reproducibility. Agentic workflows are capped at 150 iterations to bound runtime. Across 92 detections, 21 configurations, and 3 workflows, we produce \textbf{5,796 generated detections} for evaluation.

\subsection{Results}
\label{subsec:results}

To ground our automated metrics in practitioner judgment, we first validate the composite overall score against expert ratings on a 22-detection subset, then report findings across the full corpus.

\begin{tcolorbox}[rqbox, label={rq:expert-validation}]
\textbf{RQ1: Does the overall score reflect expert judgment?}
How well do automated metrics align with security practitioner ratings?
\end{tcolorbox}

Because obtaining expert ratings is time-consuming and costly, we validate 
our metrics on a held-out subset of 22 detections. For each generated 
detection, multiple security practitioners rated the effort required to 
transform the artifact into the gold-standard detection on a 0--10 scale 
(0 = substantial effort; 10 = ready to deploy). We averaged the expert 
ratings and compared them against the automated composite score.

\paragraph{Weight Calibration.}
The composite overall score combines semantic and syntactic similarity with 
configurable weights. To select appropriate weights, we evaluated 
weightings from 0.0 to 1.0 in 0.2 increments (Table~\ref{tab:weight-sensitivity}). 
The 60/40 semantic-syntactic weighting maximizes Spearman correlation 
($\rho = 0.65$), whereas 80/20 matches the lowest mean absolute error 
(MAE = 0.11 vs.\ 0.16) while retaining syntactic signal. Given the marginal difference in rank correlation 
($\Delta\rho = 0.01$, not statistically significant at $n = 22$) and the 
domain consideration that functional correctness should dominate 
surface-level similarity in detection quality, we select the 80/20 
weighting.

\paragraph{Validation Results.}
With the 80/20 weighting, the overall score shows strong, statistically 
significant correlation with expert ratings (Spearman $\rho = 0.64$, 
$p < 0.002$; Pearson $r = 0.63$, $p < 0.002$), with a mean absolute 
error of 0.11 on a normalized 0--1 scale. Detections that score highly 
under the automated composite tend to require minimal expert effort, 
while lower composite scores correspond to artifacts needing substantial 
revision.

\begin{table}[ht]
\centering
\caption{Weight sensitivity analysis: expert correlation by weighting.}
\label{tab:weight-sensitivity}
\begin{tabular}{cccc}
\toprule
\textbf{Semantic} & \textbf{Syntactic} & \textbf{Spearman $\rho$} & \textbf{MAE} \\
\midrule
0.0 & 1.0 & 0.35 & 0.37 \\
0.2 & 0.8 & 0.60 & 0.29 \\
0.4 & 0.6 & 0.58 & 0.22 \\
0.6 & 0.4 & \textbf{0.65} & 0.16 \\
\textbf{0.8} & \textbf{0.2} & 0.64 & \textbf{0.11} \\
1.0 & 0.0 & 0.62 & \textbf{0.11} \\
\bottomrule
\end{tabular}
\end{table}

While the weight selection and validation use the same expert sample, 
the stability of correlations across the 0.6--0.8 range 
($\rho = 0.64$--$0.65$) suggests the result is not an artifact of 
overfitting to this particular weighting. This agreement justifies 
using the composite overall metric as a scalable proxy for human evaluation 
in the remaining research questions.

\begin{tcolorbox}[rqbox, label={rq:workflow-comparison}]
\textbf{RQ2: How do authoring workflows compare?}
What are the performance differences among Baseline, Sequential, and Agentic approaches?
\end{tcolorbox}

\begin{table}[ht]
\centering
\caption{Overall score by authoring approach.}
\label{tab:similarity-by-approach}
\begin{tabular}{lccc}
\toprule
\textbf{Approach} & \textbf{Mean} & \textbf{Median} & \textbf{Std Dev} \\
\midrule
Agentic    & 0.447 & 0.419 & 0.180 \\
Sequential & 0.388 & 0.356 & 0.150 \\
Baseline   & 0.375 & 0.328 & 0.160 \\
\bottomrule
\end{tabular}
\end{table}

Table~\ref{tab:similarity-by-approach} reports overall scores across the three authoring approaches, averaged over all model configurations. The Agentic workflow yields the highest similarity (0.447 mean), outperforming both Sequential (RAG-augmented) and Baseline (zero-shot) approaches by 15\% and 19\% respectively. Sequential provides a modest gain over Baseline (+3.5\%), indicating that retrieval-augmented generation improves alignment even without full agentic reasoning. Median similarity follows the same pattern, confirming that gains are not driven by outliers.

We observe greater score variability under the Agentic approach. We attribute this to the multi-turn, tool-orchestrated nature of agentic workflows and the inherent nondeterminism of LLM generations, which can amplify dispersion across diverse detection tasks and schemas.

\begin{tcolorbox}[rqbox, label={rq:criteria-analysis}]
\textbf{RQ3: What are the strengths and weaknesses by evaluation criterion?}
Which LLM-as-a-Judge criteria have the highest and lowest pass rates, and how does tool-assisted retrieval affect them?
\end{tcolorbox}

\begin{table}[ht]
\centering
\small
\setlength{\tabcolsep}{3.5pt} 
\renewcommand{\arraystretch}{1.05}

\begin{tabular}{l c c c}
\toprule
\textbf{Criterion} &
\textbf{All} &
\textbf{Agentic} &
\textbf{Agentic+Schema} \\
& \textbf{(\%)} & \textbf{(\%)} & \textbf{(\%)} \\
\midrule
TTP Match           & 99.4 & 99.3 (\textminus0.2) & 99.6 (+0.1) \\
Syntax Validity     & 95.9 & 95.6 (\textminus0.3) & 96.3 (+0.4) \\
Library Usage       & 61.9 & 72.7 (+10.8) & 77.8 (+15.9) \\
Data Source Correct & 31.7 & 43.4 (+11.7) & 45.7 (+13.9) \\
Indicator Alignment & 30.8 & 29.6 (\textminus1.2) & 18.8 (\textminus12.0) \\
Robustness          & 23.7 & 27.5 (+3.8) & 17.4 (\textminus6.2) \\
Logic Equivalence   & 18.4 & 18.5 (+0.2) & 9.1 (\textminus9.3) \\
Output Alignment    & 18.4 & 29.1 (+10.7) & 32.4 (+14.0) \\
Schema Accuracy     & 17.6 & 24.6 (+7.1) & 21.4 (+3.8) \\
Exclusion Parity    &  8.9 & 10.7 (+1.8) & 3.6 (\textminus5.3) \\
\bottomrule
\end{tabular}
\caption{Evaluation criteria pass rates by workflow and platform scope. Deltas computed from unrounded values.}
\label{tab:criteria-pass-rates}
\end{table}

Table~\ref{tab:criteria-pass-rates} reports pass rates for each evaluation criterion, disaggregated by workflow. The \emph{Agentic Only} column isolates runs using the ReAct-based workflow; \emph{Schema-Enabled Platforms} further restricts to the two platforms (XDR and Platform B) where schema-retrieval tools are enabled.

\textbf{Universal strengths.} \emph{TTP matching} and \emph{syntax validity} exceed 95\% across all conditions, confirming that models reliably map threat descriptions to ATT\&CK techniques and produce code judged to be syntactically valid.

\textbf{Agentic gains.} Tool-assisted retrieval yields substantial improvements on schema-sensitive criteria: \emph{library usage} (+10.8 pp), \emph{data-source selection} (+11.7 pp), and \emph{output alignment} (+10.7 pp). These gains amplify further when schema tools are available, with schema-enabled platforms reaching +15.9 pp on library usage and +14.0 pp on output alignment.

\textbf{Persistent challenges.} \emph{Exclusion parity} remains the weakest criterion (<11\%), indicating that models rarely replicate the precise exclusion logic of reference detections. \emph{Logic equivalence} and \emph{indicator alignment} also decline in the schema-enabled subset, suggesting that richer schema context can bias models toward platform idioms at the expense of exact semantic fidelity to the reference.

These patterns confirm that agentic retrieval most benefits criteria requiring platform-specific knowledge (schemas, tables, libraries), while criteria dependent on preserving reference semantics (exclusions, logic structure) require additional grounding strategies.

\begin{tcolorbox}[rqbox, label={rq:reasoning-effort}]
\textbf{RQ4: What is the effect of reasoning effort?}
How do high, medium, and low reasoning tiers compare?
\end{tcolorbox}

\begin{table}[ht]
\centering
\caption{Overall similarity by reasoning tier.}
\label{tab:reasoning-tier-summary}
\begin{tabular}{lcccc}
\toprule
\textbf{Reasoning} & \textbf{All} & \textbf{All} & \textbf{Agentic} & \textbf{Agentic} \\
\textbf{Tier} & \textbf{Mean} & \textbf{Median} & \textbf{Mean} & \textbf{Median} \\
\midrule
High   & 0.442 & 0.408 & 0.501 & 0.480 \\
Medium & 0.420 & 0.372 & 0.473 & 0.437 \\
Low    & 0.410 & 0.367 & 0.438 & 0.394 \\
\bottomrule
\end{tabular}
\end{table}

Table~\ref{tab:reasoning-tier-summary} reports overall similarity by reasoning tier, disaggregated by workflow. Across all approaches, the \emph{high} tier achieves 0.442 mean similarity, with diminishing returns at lower tiers—the gap narrows from high→medium (0.022) to medium→low (0.010).

The Agentic workflow amplifies reasoning tier effects substantially. High-tier Agentic runs reach 0.501 mean similarity—a 13\% improvement over the all-approaches high tier (0.442). The tier separation also widens: high→medium spans 0.028 and medium→low spans 0.035, indicating that reasoning effort compounds with tool-assisted retrieval.

The \emph{medium} tier offers a favorable quality–cost trade-off for both workflows: it captures most high-tier gains while incurring lower token consumption and latency. For Agentic deployments, medium-tier reasoning achieves 94\% of high-tier performance (0.473 vs. 0.501), making it a practical default where throughput requirements constrain reasoning budget.

\begin{tcolorbox}[rqbox, label={rq:top-n-leaderboard}]
\textbf{RQ5: Which configurations achieve top performance?}
What model–approach combinations occupy the Top-5 positions?
\end{tcolorbox}

\begin{table}[ht]
\centering
\caption{Top-5 leaderboard by model, reasoning tier, and approach.}
\label{tab:top-n-leaderboard}
\begin{tabular}{rlllcc}
\toprule
\textbf{Rank} & \textbf{Model} & \textbf{Tier} & \textbf{Approach} & \textbf{Mean} & \textbf{Median} \\
\midrule
1 & gpt-5      & medium & Agentic & 0.578 & 0.545 \\
2 & gpt-5.1    & high   & Agentic & 0.575 & 0.549 \\
3 & gpt-5-mini & high   & Agentic & 0.553 & 0.541 \\
4 & gpt-5.1    & medium & Agentic & 0.547 & 0.520 \\
5 & gpt-5      & high   & Agentic & 0.545 & 0.526 \\
\bottomrule
\end{tabular}
\end{table}

Table~\ref{tab:top-n-leaderboard} reports the top five performing configurations. All five are Agentic workflows, with the GPT-5 family claiming three of the top five positions. Notably, medium reasoning tier appears in both \#1 and \#4 positions, while high tier occupies the remaining slots—reinforcing that medium offers competitive quality at lower cost. While high reasoning generally outperforms medium on aggregate (RQ4), the emergence of GPT-5 Medium as the top-ranking configuration (\#1) highlights a nuance in reasoning dynamics: for certain detection tasks, maximal reasoning effort can occasionally induce ``reasoning drift'' or over-complication that diverges from the gold-standard's detection logic, whereas medium effort offers an optimal balance. The tight clustering of scores (0.545–0.578) across GPT-5 and GPT-5.1 families indicates that model selection within the frontier tier matters less than workflow choice (Agentic) and reasoning configuration.

Crucially, these results are achieved \emph{without executing the generated code, iterative refinement via execution feedback, or automated validation}. The current MCP tool suite lacks efficacy tools—such as query execution against sample telemetry or unit test runners—that would enable agents to self-correct. Generated detections therefore represent \emph{first-draft artifacts} that detection authors refine, rather than production-ready rules. The scores reported here reflect alignment with gold-standard detections at the authoring stage; extending the tool suite with execution-based feedback loops represents a natural next step to close the gap between draft and deployable code.

\begin{tcolorbox}[rqbox, label={rq:token-efficiency}]
\textbf{RQ6: What are the cost–quality trade-offs?} How do workflows compare in token consumption?
\end{tcolorbox}

The Agentic workflow consumes roughly two orders of magnitude more tokens than Baseline (using median token counts: \textasciitilde80$\times$ Baseline), reflecting the multi-turn ReAct pattern~\cite{yao2023reactsynergizingreasoningacting} and repeated tool invocations. Sequential occupies a middle ground---about 2$\times$ Baseline at the median---offering retrieval-augmented generation with predictable cost. Token variance is highest for Agentic, exhibiting a heavy-tailed distribution as planning depth varies substantially with detection complexity.

Sequential achieves 87\% of Agentic quality (RQ2) at \textasciitilde40$\times$ lower median token cost, making it suitable for routine detection authoring where throughput and cost efficiency are priorities. However, reaching the top-tier performance observed in RQ5---where the best configurations exceed 0.57 score---\emph{requires} the Agentic workflow. For complex detections involving unfamiliar schemas, novel threat patterns, or multi-stage logic, the added quality premium and access to iterative tool-assisted refinement justify the resource investment.

\begin{tcolorbox}[rqbox, label={rq:platform-language}]
\textbf{RQ7: How do scores vary across platforms and languages?}
Does detection complexity explain performance differences?
\end{tcolorbox}

\begin{table}[ht]
\centering
\caption{Overall similarity by platform.}
\label{tab:platform-performance}
\begin{tabular}{l c c c}
\toprule
\textbf{Platform} & \textbf{All}  & \textbf{Agentic} & \textbf{Avg LoC} \\
\midrule
Internal Platform A  & 0.488 & 0.511 (+0.023) & 60 \\
Sentinel             & 0.458 & 0.500 (+0.042) & 93 \\
XDR                  & 0.402 & 0.411 (+0.008) & 64 \\
Internal Platform B  & 0.346 & 0.438 (+0.092) & 986 \\
Internal Platform C  & 0.322 & 0.362 (+0.040) & 139 \\
\bottomrule
\end{tabular}
\end{table}

\begin{table}[ht]
\centering
\caption{Overall similarity by language.}
\label{tab:language-performance}
\begin{tabular}{l c c c}
\toprule
\textbf{Language} & \textbf{All} & \textbf{Agentic} & \textbf{Avg LoC} \\
\midrule
Scala  & 0.488 & 0.511 (+0.023) & 60 \\
KQL    & 0.437 & 0.467 (+0.029) & 82 \\
Python & 0.334 & 0.400 (+0.066) & 562 \\
\bottomrule
\end{tabular}
\end{table}

Tables~\ref{tab:platform-performance} and~\ref{tab:language-performance} reveal variation across platforms and languages, with an inverse relationship between detection complexity and similarity scores. Internal Platform A (Scala, 60 LoC mean) and Sentinel (KQL, 93 LoC) achieve the highest scores, while Internal Platform B (Python, 986 LoC) and Internal Platform C (Python, 139 LoC) score lowest.

This pattern admits two complementary explanations. First, \emph{longer detections are harder to reproduce}: Python-based platforms encode multi-stage detection pipelines with extensive preprocessing, feature engineering, and aggregation logic that models struggle to replicate exactly. Second, \emph{simpler query languages constrain variance}: KQL and Scala detections follow more stereotyped patterns with fewer degrees of freedom, making semantic alignment easier.

Notably, the Agentic workflow provides the largest gains precisely where overall scores are lowest: Internal Platform B improves by +0.092 and Python by +0.066, compared to +0.023 for Internal Platform A. This suggests that tool-assisted retrieval disproportionately benefits complex detections, where schema lookups and similar-detection examples help models navigate platform-specific idioms that would otherwise require extensive implicit knowledge.

\begin{tcolorbox}[rqbox, label={rq:model-timeline}]
\textbf{RQ8: How does model performance evolve over time?}
Do newer model releases yield better detection quality?
\end{tcolorbox}

\begin{figure}[ht]
\centering
\includegraphics[width=\columnwidth]{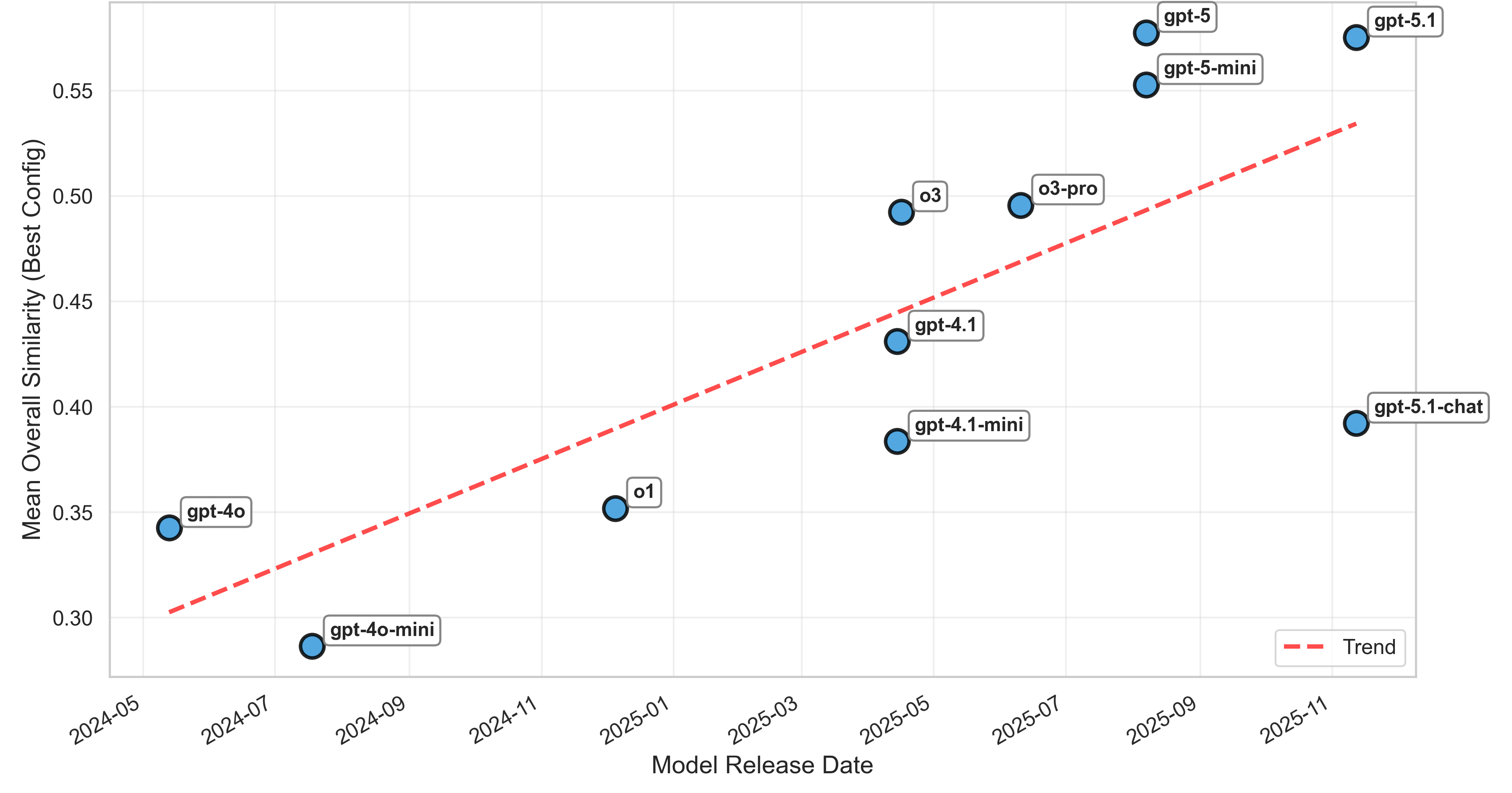}
\caption{Best-performing configuration per model over time (Agentic workflow).}
\label{fig:model-timeline}
\end{figure}

Figure~\ref{fig:model-timeline} plots the best-performing configuration for each model family under the Agentic workflow against release date. A strong positive correlation between release date and detection quality emerges ($r = 0.75$, Pearson): models released in mid-2024 (gpt-4o, gpt-4o-mini) achieve mean similarity scores below 0.35, while frontier models from late 2025 (gpt-5, gpt-5.1) exceed 0.57—a 68\% relative improvement over 18 months.

Three distinct capability tiers are visible. Early chat models (gpt-4o family) establish a baseline around 0.30. The April 2025 cohort (gpt-4.1, o3) marks a step-change, with o3-high reaching 0.49. The August–November 2025 releases (gpt-5, gpt-5.1) consolidate gains above 0.55, suggesting that detection authoring benefits from advances in both reasoning depth and instruction following.

Notably, GPT-5.1 (0.575) does not surpass GPT-5 (0.578) despite being released three months later. This aligns with observations in other works~\cite{benchek2025aiconsumerindexace, dutta2026empiricalinvestigationrobustnesslarge}, where GPT-5.1 similarly trails GPT-5 by a small margin. OpenAI has indicated that GPT-5.1 prioritizes inference speed and cost efficiency over raw capability gains, representing a trade-off between quality and operational efficiency rather than a strict successor~\cite{oai_gpt5_5.1}.

\section{Discussion}
\label{sec:discussion}

This section reflects on the strengths and limitations of our approach, interprets the impact of schema availability on detection quality, and outlines directions for future work.

\textbf{The Impact of Schema Availability.} A critical component of our evaluation was the selective enabling of MCP schema tools for specific platforms (XDR and Platform B). This experimental condition allows us to isolate the impact of retrieval-augmented context on generation quality. The results demonstrate a sharp distinction between ``structural'' and ``logical'' correctness. For the schema-enabled platforms, metrics that rely on exact knowledge retrieval showed significant gains over the all-approach average: \textbf{Library Usage} improved by 15.9 percentage points (77.8\% vs 61.9\%), \textbf{Data Source Correctness} by 14.0 percentage points (45.7\% vs 31.7\%), and \textbf{Output Alignment} by 14.0 percentage points (32.4\% vs 18.4\%). This validates the core premise of \name{}: providing agents with tool-based access to schemas directly mitigates hallucinations in table selection. However, \textbf{Schema Accuracy} (21.4\%) remained low even with tools available. This suggests that while agents successfully identify the broad components (correct table, correct libraries), they still struggle with the fine-grained precision of specific field names within complex, production-grade schemas—likely due to context window constraints or retrieval precision limits.

\textbf{The ``Tribal Knowledge'' Gap.} While structural metrics improved with tools, semantic metrics like \textbf{Exclusion Parity} (3.6\%) and \textbf{Logic Equivalence} (9.1\%) remained the most challenging criteria, particularly in the schema-aware subset. This highlights a fundamental limitation: detection logic often relies on ``tribal knowledge''—unwritten environmental context (e.g., ``server X is a benign scanner'')—rather than documented schemas. The availability of a database schema does not help an agent infer that a specific IP range should be excluded from a detection. This finding reinforces that AI can act as a powerful accelerator for drafting the \textit{skeleton} and \textit{syntax} of a detection, but human engineers remain essential for injecting environment-specific context and robust exclusion logic.

\textbf{Model Selection Trade-offs.} Our evaluation reveals clear performance stratification across model families. Reasoning-capable models (GPT-5, GPT-5.1, o3-pro) consistently outperform chat/completion models, with GPT-5 achieving the top composite score (0.578) at medium reasoning effort. The medium reasoning tier offers a favorable quality--cost balance, closing most of the gap to high-effort configurations while reducing token consumption (RQ6). Notably, smaller models struggle to follow complex instructions in the Agentic workflow, where accumulated context degrades instruction adherence. Practitioners should select models based on detection complexity and latency requirements.

\textbf{Challenges and Operational Trade-offs.} Despite promising results, several challenges remain. First, LLM-driven generation exhibits variability, particularly in Agentic workflows where multi-turn nondeterminism amplifies output dispersion. Second, the Agentic approach introduces significant operational overhead. Agents accumulate large volumes of intermediate context, leading to context window exhaustion on models and significantly higher latency compared to single-shot inference. This necessitates a tiered deployment strategy: utilizing cost-effective \textbf{Sequential} workflows for standard detection drafting, while reserving resource-intensive \textbf{Agentic} workflows for complex, novel threat scenarios where the 19\% quality gain justifies the computational cost.

\textbf{Scope and Future Directions.} Our evaluation reflects deliberate design choices that balance depth with breadth, while identifying opportunities for extension. We focused on OpenAI models given their widespread adoption in enterprise settings; the MCP-based architecture is provider-agnostic by design, enabling straightforward extension to other frontier models (Anthropic, Google, Meta) as future work. Our evaluation measures \textit{authoring proximity} to gold-standard detections rather than runtime efficacy—a practical choice that enabled systematic comparison across 5,796 generated artifacts. Complementary runtime validation using synthetic attack replay would bridge this gap, measuring true positive rates and analyst triage burden in controlled environments. The 92-detection corpus spans five platforms and three languages, providing meaningful diversity in syntax, semantics, and telemetry schemas; cross-organizational validation would further establish generalizability of coding conventions and exclusion patterns. Finally, expert validation on a 22-detection subset—while constrained by practitioner availability—yielded statistically significant correlation, confirming that automated metrics reliably reflect practitioner judgment. Expanding expert coverage and incorporating feedback loops for iterative refinement represent natural extensions of this work.
\section{Conclusion}
\label{sec:conclusion}

This paper presented \name{}, an agentic framework for automating security detection authoring. \name{} leverages AI-assisted metadata extraction from organizational detection artifacts and operationalizes this knowledge through MCP-based tool orchestration. By integrating directly into developer workflows via Visual Studio Code and GitHub Copilot, \name{} enables detection authors to generate high-quality detections without disrupting established practices.

We evaluated three authoring workflows---Baseline, Sequential, and Agentic---across 92 detections, 11 models, and 21 configurations, producing 5,796 generated artifacts. Key findings include: (1) expert validation confirms strong correlation between automated metrics and practitioner judgment ($\rho = 0.64$, $p < 0.002$); (2) the Agentic workflow achieves 19\% higher overall similarity than Baseline but consumes more tokens; (3) reasoning-capable models (GPT-5, o3-pro) consistently outperform non-reasoning models; and (4) the medium reasoning tier offers a favorable quality--cost balance.

While agentic workflows demonstrate superior accuracy, the trade-offs in latency and resource consumption underscore the need for cost-aware deployment strategies. Future work will extend beyond static code generation to runtime validation, incorporating synthetic telemetry replay and controlled execution in production environments to bridge the gap between authoring quality and operational efficacy.


\bibliographystyle{ACM-Reference-Format}
\bibliography{paper}







\end{document}